\documentclass[%
twocolumn,
superscriptaddress,
amsmath,
amssymb,
aps,
prb,
floatfix,
]{revtex4-2}

\usepackage{graphicx} 

\usepackage{dcolumn}         
\usepackage{bm}              
\usepackage{hyperref}        
\usepackage{xcolor}          
\usepackage{physics}         
\usepackage{siunitx}         

\usepackage{tikz}

\usetikzlibrary{calc, arrows.meta, decorations.pathmorphing, decorations.markings}

\hypersetup{
    colorlinks = true,
    linkcolor  = blue,
    citecolor  = blue,
    urlcolor   = blue,
}

\begin{document}

\title{Weyl superconductivity from Feshbach resonance \\ in the three-dimensional repulsive Hubbard model}

\author{Taekyoung Kim}
\email{t.kim-1@tudelft.nl}
\altaffiliation[Present address: ]{QuTech, Delft University of Technology, 2600 GA Delft, The Netherlands}
\affiliation{Department of Physics and Arnold Sommerfeld Center for Theoretical Physics (ASC),
Ludwig-Maximilians-Universit\"at M\"unchen, Theresienstr. 37, M\"unchen D-80333, Germany}
\affiliation{Technical University of Munich, TUM School of Natural Sciences, Physics Department, 85748 Garching, Germany}

\author{Pit Bermes}
\affiliation{Department of Physics and Arnold Sommerfeld Center for Theoretical Physics (ASC),
Ludwig-Maximilians-Universit\"at M\"unchen, Theresienstr. 37, M\"unchen D-80333, Germany}
\affiliation{Munich Center for Quantum Science and Technology (MCQST), Schellingstr. 4, D-80799 M\"unchen, Germany}

\author{Giorgia A. Busin}
\affiliation{Department of Physics and Arnold Sommerfeld Center for Theoretical Physics (ASC),
Ludwig-Maximilians-Universit\"at M\"unchen, Theresienstr. 37, M\"unchen D-80333, Germany}
\affiliation{Munich Center for Quantum Science and Technology (MCQST), Schellingstr. 4, D-80799 M\"unchen, Germany}

\author{Fabian Grusdt}
\email{Fabian.Grusdt@lmu.de}
\affiliation{Department of Physics and Arnold Sommerfeld Center for Theoretical Physics (ASC),
Ludwig-Maximilians-Universit\"at M\"unchen, Theresienstr. 37, M\"unchen D-80333, Germany}
\affiliation{Munich Center for Quantum Science and Technology (MCQST), Schellingstr. 4, D-80799 M\"unchen, Germany}

\date{\today}

\begin{abstract}
Motivated by the first experimental realization of the antiferromagnetic phase transition in the three-dimensional Fermi-Hubbard model, we present a theoretical study of the model's 3D superconducting phase. By formulating a 3D extension of the Feshbach mechanism, we provide a unified microscopic picture at strong coupling in which pairing is driven by near-resonant bound states of dopants. These long-lived bound states acquire a qualitatively different internal structure in three dimensions compared to their two-dimensional counterparts, giving rise to a distinct superconducting state, namely a time-reversal symmetry breaking $d_{x^2-y^2}+id_{z^2}$ pairing state. We also provide an estimate of the corresponding critical temperature, and provide evidence that the resulting superconducting phase hosts gapless Weyl points. Our results represent a new milestone for the field of quantum simulation, challenging experiments and large-scale numerics alike to test our predictions.
\end{abstract}

\maketitle

\section{\label{sec:intro}Introduction}

The historical trajectory of many-body physics demonstrates that highly simplified, minimal models are often more effective at capturing the universal properties of complex phenomena than exhaustive microscopic descriptions. When studied in great detail, even the conceptually simplest models provide new benchmarks for theorists and experimentalists to test new hypotheses~\cite{niss2009, niss2011}. In this vein, the Fermi-Hubbard model requires rigorous and exhaustive investigation. It is regarded as one of the most paradigmatic many-particle models one can conceive, that cannot be reduced to a single-particle description~\cite{auerbach1994}. Widely regarded as a key concept in understanding high-temperature superconductivity, the Fermi-Hubbard model describes a wide range of physical phenomena resulting from strong electron–electron correlations~\cite{arovas2022}.

The ground state correlations and excitations of the Fermi-Hubbard model in one dimension have been understood employing several innovative strategies, including the Bethe ansatz, bosonization, the Luttinger and Tomonaga models, and the perturbative renormalization group~\cite{giamarchi2003}.
Unfortunately, these methods are specialized to one dimension and their extension to two or three dimensions is difficult and has not been achieved~\cite{auerbach1994}. At the same time, numerical approaches such as density-matrix renormalization group (DMRG) become prohibitively costly in higher dimensions.

\begin{figure}[t] 
    \centering
    \includegraphics[width=1.0\columnwidth]{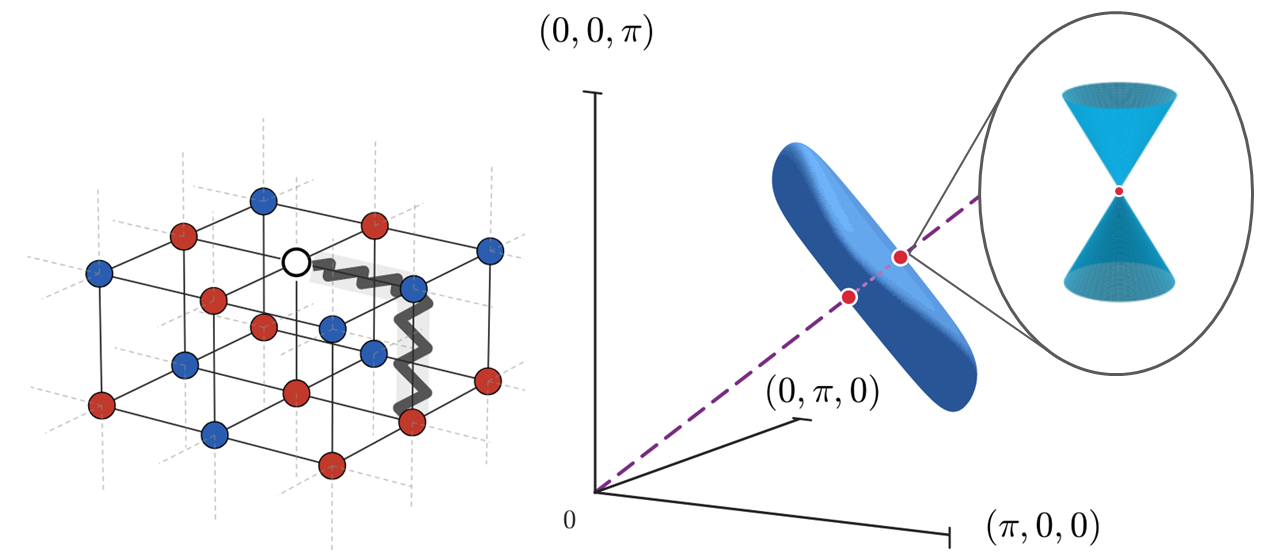}
    \caption{(Left) Magnetic polaron in a 3D cubic lattice with antiferromagnetic ordering. Sites occupied by spin-up electrons are indicated in red, while those with spin-down electrons are indicated in blue. The dopant creates a string of displaced spins $\Sigma$ (shown in gray) as it moves through the lattice, confining it in a potential. (Right) Fermi pockets of the magnetic polarons, shown in blue, within one octant of the crystal Brillouin zone. The dashed line denotes the nodal line of the $d_{x^2-y^2}+id_{z^2}$ superconducting gap we predict. Its intersection with the Fermi pockets produces two Weyl points per octant, around which the quasiparticle spectrum disperses linearly.}
    \label{fig:overview}
\end{figure}

The superconducting phase of the two-dimensional Fermi-Hubbard model constitutes an ongoing topic of research~\cite{arovas2022}. While some numerical studies indicate that the vanilla Fermi-Hubbard model at typical parameters does not exhibit superconductivity in the ground state~\cite{qin2020}, a superconducting phase has been predicted to emerge with the inclusion of next-to-nearest neighbor hopping terms~\cite{xu2024}. While the exact ground state remains unknown, the Fermi-Hubbard model is believed to host a superconducting $d_{x^2-y^2}$-wave pairing state, as supported by studies using the functional renormalization group ~\cite{vilardi2019}. On a square lattice, $d_{x^2-y^2}$-wave symmetry requires the superconducting gap to have nodes along the Brillouin zone diagonals.

The superconducting phase of the three-dimensional (3D) Fermi-Hubbard model, on the other hand, is largely unexplored. Early theoretical works focused on a weak-coupling limit, indicating $d$-wave pairing~\cite{scalapino1986}. To bypass computational limitations, physicists have turned attention to quantum simulators with ultracold fermions~\cite{esslinger2010, tarruell2018, bohrdt2021}. Recently, researchers used this technology to successfully observe an antiferromagnetic phase transition in a uniform 3D system~\cite{shao2024}. This major breakthrough motivates future research to focus on realizing the superconducting phase of the repulsive 3D Hubbard model, mapping out its phase diagram and characterizing the symmetry of the superconducting state.

The underlying pairing mechanism responsible for superconductivity in two dimensions has been extensively investigated~\cite{keimer2015, manousakis1991, chen2024}, with a recent proposal suggesting a Feshbach pairing mechanism~\cite{homeier2024, homeier2025}. The Feshbach pairing mechanism is based on magnetic polarons~\cite{kane1989, sachdev1989, martinez1991} and bipolarons~\cite{grusdt2023, bohrdt2023} realizing an emergent Feshbach resonance. It is theorized that magnetic polarons describe the fermionic quasiparticles formed by doped holes in a Mott insulator. The polarons form a Fermi liquid at low doping, and they interact weakly with each other by combining to form long-lived bound states known as bipolarons, enabling an emergent Feshbach resonance. Motivated by Cuprate materials, whose relevant physics can be modelled in terms of two dimensions, the properties of magnetic polarons and the Feshbach mechanism have been studied so far in antiferromagnetic systems in two dimensions~\cite{homeier2024, homeier2025, bermes2024, grusdt2023}. However, the Feshbach formalism can be extended to higher dimensions, the topic of this work.

Motivated by the first experimental realization of an antiferromagnetic phase transition in the 3D Hubbard model~\cite{shao2024}, in this paper we investigate unconventional superconductivity in three dimensions. We extend and establish the previously proposed Feshbach pairing mechanism to three dimensions. Our main contribution lies in (i) deducing the 3D superconducting state to have $d+i d$~-wave symmetry, realizing a Weyl superconductor~\cite{meng2012} characterized by isolated gapless nodal points known as Weyl points, and in (ii) providing estimates of the corresponding critical temperature $T_c$. Within this work, we consider a cubic lattice (see Fig.~\ref{fig:overview}) having octahedral symmetry. Beyond the extension of the strong-coupling Feshbach mechanism from two to three dimensions, we complement our analysis by furnishing microscopic expressions of the corresponding Ginzburg-Landau formalism.
Tests of our predictions, by large-scale numerical simulations or quantum simulators, constitute a new milestone for these fields to demonstrate their capabilities.

\section{\label{sec:formalism}Formalism}

We begin from a strong-coupling theory of spin-$\frac{1}{2}$ fermions. Assuming a ground state with long-range antiferromagnetic order, holes doped into this background form magnetic polarons that undergo a Feshbach resonance, driving a superconducting instability. Within the BEC-BCS crossover framework, we focus on the BCS side.

\subsection{Model}

Our formalism is based on the 3D cubic lattice $t\text{-}J$ model, which can be obtained from the Fermi-Hubbard model in the limit of strong on-site interactions~\cite{auerbach1994}, as a description of a doped antiferromagnet.
\begin{equation}
\begin{aligned}
\hat{\mathcal{H}}_{t-J}
= {} & \hat{\mathcal{H}}_{t}+\hat{\mathcal{H}}_{J}\\
= {} & -t \sum_{\langle \mathbf{i},\mathbf{j} \rangle, \sigma}
\hat{\mathcal{P}}\!\left( \hat{c}^\dagger_{\mathbf{i},\sigma}\hat{c}_{\mathbf{j},\sigma} + \text{H.c.} \right)\!\hat{\mathcal{P}}
\\
& {} + J_z \sum_{\langle \mathbf{i},\mathbf{j} \rangle}
\left( \hat{S}^z_{\mathbf{i}} \hat{S}^z_{\mathbf{j}} - \frac{1}{4}\hat{n}_{\mathbf{i}} \hat{n}_{\mathbf{j}} \right)
\\
& {} + \frac{J_\perp}{2} \sum_{\langle \mathbf{i},\mathbf{j} \rangle}
\left( \hat{S}^+_{\mathbf{i}} \hat{S}^-_{\mathbf{j}} + \text{h.c.} \right).
\end{aligned}
\label{eq:hamiltonian}
\end{equation}

The strength of the nearest-neighbor hopping term is $t$, and the operators $\hat{c}^\dagger_{\mathbf{i},\sigma}$ and $\hat{c}_{\mathbf{j},\sigma}$ are the fermion creation and annihilation operators with spin $\sigma$ on site $\mathbf{i}$, and $\hat{n}_{\mathbf{i}}$ is the number operator on site $\mathbf{i}$. The Gutzwiller projector $\hat{\mathcal{P}}$ prevents double occupancies, and the last two terms represent the antiferromagnetic interaction. Here we assume $J=J_z=J_\perp$ and focus on the case $t/J=3$. At half-filling and low temperatures, the fermions each occupy a single site, and the lattice has antiferromagnetic (AFM) ordering~\cite{auerbach1994}.

\subsection{\label{sec:feshbach}Two-channel model and Feshbach hypothesis}

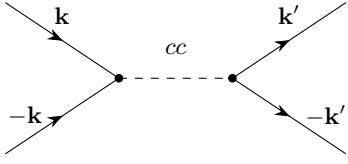
\begin{figure}[t]
\centering
\begin{tikzpicture}
  \coordinate (v1) at (1.5,0);      
  \coordinate (v2) at (3,0);      
  \coordinate (in1) at (0,1);     
  \coordinate (in2) at (0,-1);    
  \coordinate (out1) at (4.5,1);    
  \coordinate (out2) at (4.5,-1);   

  \draw (in1) -- node[midway, above =1mm] {$\mathbf{k}$} (v1);
  \draw[-{Stealth[length=2mm]}] ($(in1)!0.5!(v1)$) -- ++(0.01,-0.005); 

  \draw (in2) -- node[midway, left=1.5mm] {$-\mathbf{k}$} (v1);
  \draw[-{Stealth[length=2mm]}] ($(in2)!0.5!(v1)$) -- ++(0.01,0.005); 

  \draw[dashed] 
    (v1) -- node[midway, above=2mm] {$cc$} (v2);

  \draw (v2) -- node[midway, above =1mm] {$\mathbf{k}'$} (out1);
  \draw[-{Stealth[length=2mm]}] ($(v2)!0.5!(out1)$) -- ++(0.01,0.005); 

  \draw (v2) -- node[midway, right=1mm] {$-\mathbf{k}'$} (out2);
  \draw[-{Stealth[length=2mm]}] ($(v2)!0.5!(out2)$) -- ++(0.01,-0.005); 

  \node[draw, circle, inner sep=1.0pt, fill=black] at (v1) {};
  \node[draw, circle, inner sep=1.0pt, fill=black] at (v2) {};
  
\end{tikzpicture}
\caption{Two-body interactions underlying the two-channel model of the Feshbach hypothesis. The filled lines represent propagators of polarons with momenta ($\mathbf{k}$, $-\mathbf{k}$) and ($\mathbf{k}'$, $-\mathbf{k}'$), and the dotted line represents the bipolaron ($\textit{cc}$), which acts as a virtual long-lived bound state.}
\label{fig:feshbach}
\end{figure}

In this section we will outline the pairing mechanism known as the Feshbach hypothesis of high temperature superconductivity~\cite{homeier2024, homeier2025, bermes2026}. We restrict our attention to the singlet pairing channel. The Feshbach hypothesis theorizes that two polarons pair by combining to form a virtual bound state (a bipolaron) rather than exchange a boson carrying some momentum. A diagram of this process is shown in Fig \ref{fig:feshbach}. The term in the Hamiltonian describing this coupling is given as follows,
\begin{equation}
    \hat{\mathcal{H}}_{\text{int}} = \sum_{\alpha=1}^{n} \mathcal{M}_{\alpha}(\mathbf{k}) \,\hat{b}_{\alpha}^{\dagger}\,\hat{\pi}_{\mathbf{k},\downarrow}\,\hat{\pi}_{-\mathbf{k},\uparrow} + \text{h.c.}
    \label{eq:int}
\end{equation}
The operators $\hat{\pi}_{-\mathbf{k}, \uparrow} \hat{\pi}_{\mathbf{k}, \downarrow}$ annihilate a pair of polarons at opposite momenta, while $\hat{b}_{\mathbf{\alpha}}^\dagger$ creates a bipolaron in the ground state at zero momentum. If this state is $n-$fold degenerate, $(\alpha=1,\ldots,n)$ labels the individual states. The scattering matrices $\mathcal{M}_{\alpha}(\mathbf{k})$ are defined by
\begin{equation}
    \mathcal{M}_{\alpha} (\mathbf{k}) = \langle 0 | \hat{b}_{\alpha} \hat{\mathcal{H}}_{\perp} \hat{\pi}_{-\mathbf{k}, \uparrow}^\dagger \hat{\pi}_{\mathbf{k}, \downarrow}^\dagger | 0 \rangle.
    \label{eq:Mform}
\end{equation}
Here $\hat{\mathcal{H}}_{\perp}$ denotes the part of the full $t\text{-}J$ Hamiltonian in Eq.~\ref{eq:hamiltonian} that couples these two channels, which is the spin flip term $J_\perp$. The state $| 0 \rangle$ represents the AFM background, which plays the role of the vacuum. The spin flip term acting on a pair of polarons generates a finite matrix element with the bipolaron state by combining the strings associated with the two dopants into a single string connecting them~\cite{homeier2024}.

Integrating out the bipolaron states in Eq.~\ref{eq:int} using the Schrieffer-Wolff transformation~\cite{coleman2015} yields the following many-body Hamiltonian,  
\begin{equation}
\begin{split}
\hat{\mathcal{H}}_{\text{eff}} &= \sum_{\mathbf{k},\sigma} (\epsilon_{\mathbf{k}}^{\text{sc}} - \mu)\,\hat{\pi}_{\mathbf{k},\sigma}^{\dagger}\,\hat{\pi}_{\mathbf{k},\sigma} \\
&\quad + \sum_{\mathbf{k},\mathbf{k}'} V_{\mathbf{k},\mathbf{k}'} \,\hat{\pi}_{-\mathbf{k},\uparrow}^{\dagger}\,\hat{\pi}_{\mathbf{k},\downarrow}^{\dagger}\,\hat{\pi}_{-\mathbf{k}',\uparrow}\,\hat{\pi}_{\mathbf{k}',\downarrow},
\end{split}
\end{equation}
where the two-body interaction $V_{\mathbf{k}, \mathbf{k}'}$ is given by
\begin{equation} \label{eq:potential_expanded}
    V_{\mathbf{k},\mathbf{k}'} = \frac{1}{2V} \sum_{\alpha=1}^{n} \mathcal{M}_{\alpha}(\mathbf{k})\mathcal{M}_{\alpha}(\mathbf{k}') \left[ \frac{1}{2\epsilon_{\mathbf{k}}^{\text{sc}}-E_0} + \frac{1}{2\epsilon_{\mathbf{k'}}^{\text{sc}}-E_0} \right].
\end{equation}
Here, $E_0$ denotes the lowest bipolaron energy, $\epsilon_{\mathbf{k}}^{\text{sc}}(\mu)$ is the magnetic polaron dispersion (chemical potential), and $V$ represents the volume of the system. The functions $\mathcal{M}_{\alpha}(\mathbf{k})$ can always be chosen to be real valued~\cite{homeier2024}.

\section{\label{sec:results}Constituting charge carriers: magnetic polarons and bipolarons}

At low doping, magnetic polarons form a Fermi liquid whose Fermi surface geometry is determined by the single polaron dispersion, and they can recombine into a meta-stable bipolaron state at higher energies. In this section we derive the properties of those constituents in the 3D Hubbard model.

\subsection{Truncated basis}

\subsubsection{Single dopant}\label{sec:truncsi}

We construct an effective Hamiltonian in a truncated Hilbert space, following Ref.~\cite{bermes2024}. This method has been shown to capture all qualitative features of the magnetic polaron in two dimensions very well and efficiently~\cite{bermes2024}. We approximate the ground state at half filling as a classical N\'eel state. This treatment is well justified in the strong coupling regime ($t\gg J$), where the hopping of dopants occurs on a much faster timescale than the spin fluctuations. The movement of the dopants therefore creates a string of displaced spins $\Sigma$, confining the dopant in an approximately linear potential (see Fig.~\ref{fig:overview}). The string of displaced spins $\Sigma$ together with the dopant position $\mathbf{x}$ serves as a set of labels for constructing a basis $\{\,|\mathbf{x}, \Sigma\rangle\,\}
$ of the Hilbert space. A Krylov basis is constructed by applying the hopping term of the Hamiltonian $\hat{\mathcal{H}}_{t}$ to a classical state with string length of zero, creating a truncated Hilbert space after restricting the basis to states with string lengths~$\leq l_{\mathrm{max}}$. Within this basis all the matrix elements of $\hat{\mathcal{H}}_{t}+\hat{\mathcal{H}}_{J}$ are subsequently included, and are given by
\begin{equation}
    \hat{\mathcal{H}}_{\mathrm{eff}}
    = \sum_{\mathbf{x},\Sigma}
      \sum_{\mathbf{x}',\Sigma'}
      |\mathbf{x}',\Sigma'\rangle
      \langle \mathbf{x}',\Sigma'|
      \hat{\mathcal{H}}_{t\!-\!J}
      |\mathbf{x},\Sigma\rangle
      \langle \mathbf{x},\Sigma|.
      \label{eq:effham}
\end{equation}

The diagonal elements of the Hamiltonian in this basis are the energy contributions of the string, and the off-diagonal elements are the $t$ and $J$ terms which couples different configurations of strings. Using translational invariance, we apply a Lee-Low-Pines transformation~\cite{Lee1953} to block diagonalize the Hamiltonian in momentum space~\cite{bermes2024} (see Appendix~\ref{app:LLP}). This yields the dispersion relation $\epsilon_{\mathbf{k}}^{\text{sc}}$ for a single dopant in an AFM lattice. Importantly, the dopant has kinetics enabled by the spin-flip terms and the Trugman loops, which are one and a half loops of the hole encircling a square plaquette~\cite{bermes2024}. 

\begin{figure}[t]
    \centering
    
    \begin{minipage}[t]{0.58\linewidth}
        \raggedright (a)
        \includegraphics[width=\linewidth]{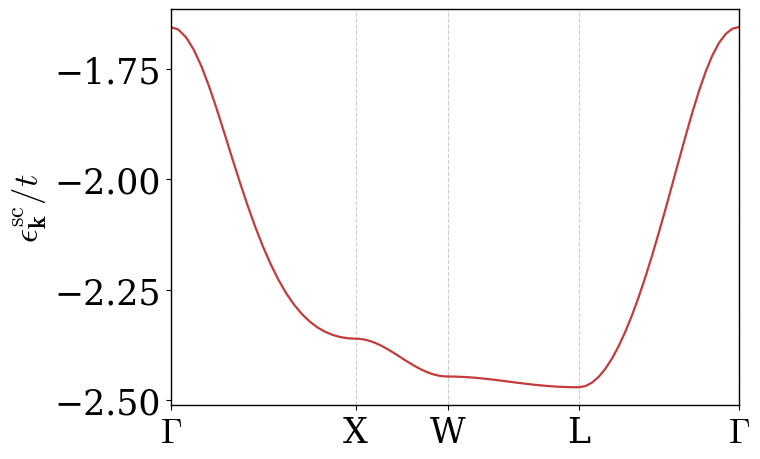}
    \end{minipage}\hfill
    \begin{minipage}[t]{0.38\linewidth}
        \raggedright (b)
        \includegraphics[width=\linewidth]{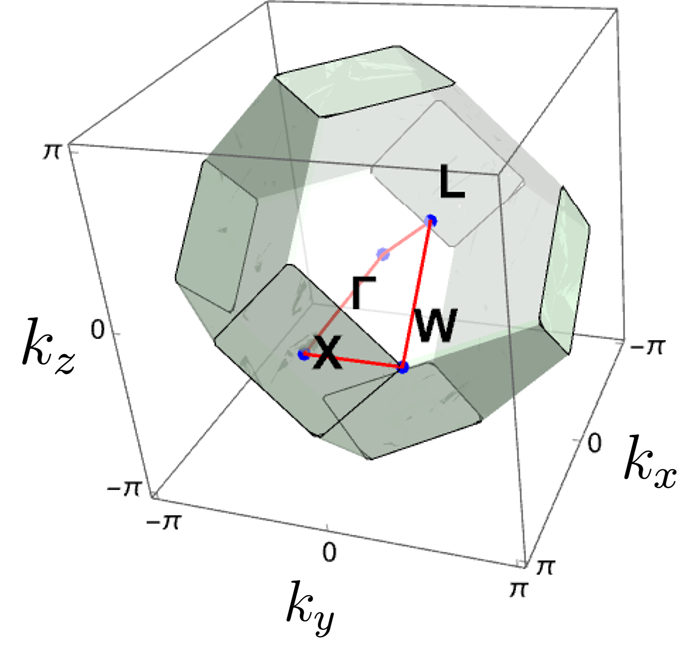}
    \end{minipage}

    \caption{(a) Polaron dispersion relation $\epsilon_{\mathbf{k}}^{\text{sc}}$ calculated from the truncated basis approach, Eq.~\ref{eq:effham} along the high-symmetry points in the magnetic Brillouin zone with $l_{\mathrm{max}}=7$. (b) Path along the high-symmetry points in red. The CBZ is shown by the outer clear cube, while the MBZ is shown in green. The path in the Brillouin zone on the left figure is $\Gamma-\mathrm{X}-\mathrm{W}-\mathrm{L}-\Gamma$, with $\Gamma=(0,0,0), \mathrm{X}=(\pi, 0, 0), \mathrm{W}=(\pi, \pi/2,0),$ and $\mathrm{L}=(\pi/2, \pi/2, \pi/2)$.}
    \label{fig:singlehole_combined_aligned}
\end{figure}

\subsubsection{Two dopants} \label{sec:truncbi}

Analogous to the single dopant case, we construct a basis for two dopants by initializing them on adjacent sites. Since one dopant can retrace the path of the other, the pair forms a bound state~\cite{Dimashko1993, homeier2024, grusdt2023, bohrdt2023, shraiman1988} which we refer to as a bipolaron. We label each basis element by the position of one dopant $\mathbf{x}_1$ and by the string of flipped spins connecting the two dopants $\Sigma_{cc}$, yielding a basis $\{\,|\mathbf{x}_1, \Sigma_{cc}\rangle\,\}
$. Similar to the single dopant case, we apply a cutoff at string length $l_{\mathrm{max}}$. We then use the Lee-Low-Pines transformation to move into the rest frame of one of the dopants. Afterwards, one additionally accounts for the particle statistics by antisymmetrizing the states, as shown in Ref.~\cite{grusdt2023}.

\subsection{Polarons and Fermi surface} \label{sec:pol}

Diagonalizing Eq.~\ref{eq:effham} in momentum space, we obtain the dispersion of a single polaron in a 3D AFM lattice. The lowest band of the resulting spectrum is shown in Fig.~\ref{fig:singlehole_combined_aligned}. Because the AFM order doubles the unit cell, the magnetic Brillouin zone (MBZ) occupies half of the original crystal Brillouin zone (CBZ). The dispersion $\epsilon_{\mathbf{k}}^{\text{sc}}$ exhibits a minimum at $(\pi/2, \pi/2, \pi/2)$. At low doping levels, the chemical potential $\mu$ is close to the minimum of the dispersion, such that the polarons interact weakly and form a Fermi liquid. In this regime, the Fermi surface of the polarons (see Fig.~\ref{fig:overview}) consists of four symmetry-related pockets centered around $(\pi/2, \pi/2, \pi/2)$ and equivalent points.

\subsection{Bipolaron state} \label{sec:bipol}

\begin{figure}[t] 
    \centering
    \includegraphics[width=0.95\columnwidth]{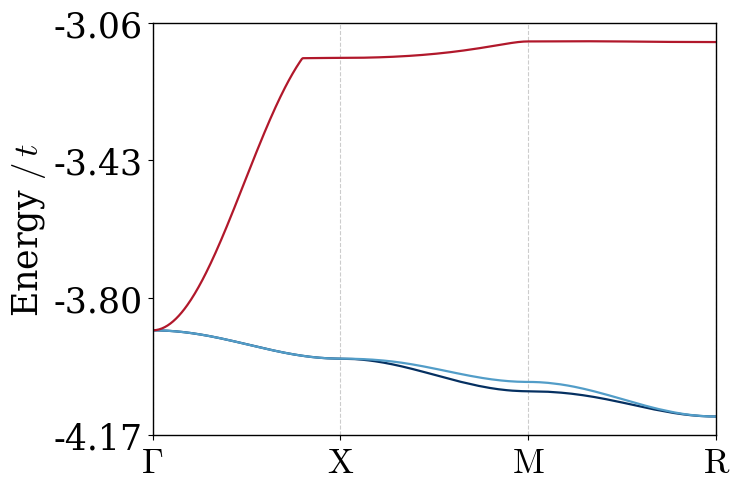}
    \caption{Bipolaron dispersion relation for the three lowest bands of the model along the path $\Gamma-\mathrm{X}-\mathrm{M}-\mathrm{R}$ in the crystal Brillouin zone, with $\Gamma=(0,0,0), \mathrm{X}=(\pi, 0, 0), \mathrm{M}=(\pi, \pi,0),$ and $\mathrm{R}=(\pi, \pi, \pi)$. Results are shown for $l_{\mathrm{max}}=7$. The model realizes its twofold degenerate ground state in $\mathrm{R}$.}
    \label{fig:bipolaron}
\end{figure}

The three lowest bipolaron bands in the AFM lattice, obtained with the truncated basis, are shown in Fig.~\ref{fig:bipolaron}. Since the bipolaron is not restricted to a sublattice, the bipolaron states are defined in the full CBZ~\cite{homeier2024}. The lowest band has a minimum at $\mathrm{R} =(\pi,\pi,\pi)$ and a maximum at $(0,0,0)$. The third band appears at a higher energy and is separated from the two lowest bands by a large gap of order $\simeq t$ around $\mathrm{R}$. At $\Gamma=(0,0,0)$ all three bands become degenerate, but with an overall energy offset $\simeq J$ above the minimum at $\mathrm{R}$.

The bipolaron momentum $Q=0$, and equivalently $\mathbf{Q}=(\pi,\pi,\pi)$ in the MBZ, is relevant for the Feshbach mechanism discussed in Sec.~\ref{sec:feshbach}. We focus on the first possible Feshbach resonance, where the lowest-energy bipolaron state is energetically just above the magnetic polaron continuum. Tuning parameters, such as nearest neighbor interaction or next-to-nearest neighbor tunneling may be used to move the system across resonance, as in the two-dimensional case~\cite{blatz2026}. Determining the precise location of the resonance will require large scale numerical simulations.

At the rotational invariant momentum $(\pi,\pi,\pi)$, the twofold degenerate bipolaron ground states $\ket{\psi_{\alpha}}$ ($\alpha = 1,2$) transform as eigenstates of the rotational operators $\hat{R}$ of the octahedral symmetry group, satisfying  $\hat{R}\,\ket{\psi_{\alpha}}=\lambda_\alpha(\hat{R})\,\ket{\psi_{\alpha}}$. The eigenvalues $\lambda_\alpha(\hat{R})$ characterize the action of the rotation on the degenerate subspace. From their values, we can infer that the degenerate ground states $\ket{\psi_{\alpha}}$ furnish the two-dimensional irreducible representation $E$ of the octahedral group. Consequently, the ground state's degeneracy is protected by the symmetry of the lattice, and lowering the symmetry will break it. More supporting evidence is shown in Appendix~\ref{app:group}.

\begin{figure}[t]
    \centering
    \includegraphics[width=\columnwidth]{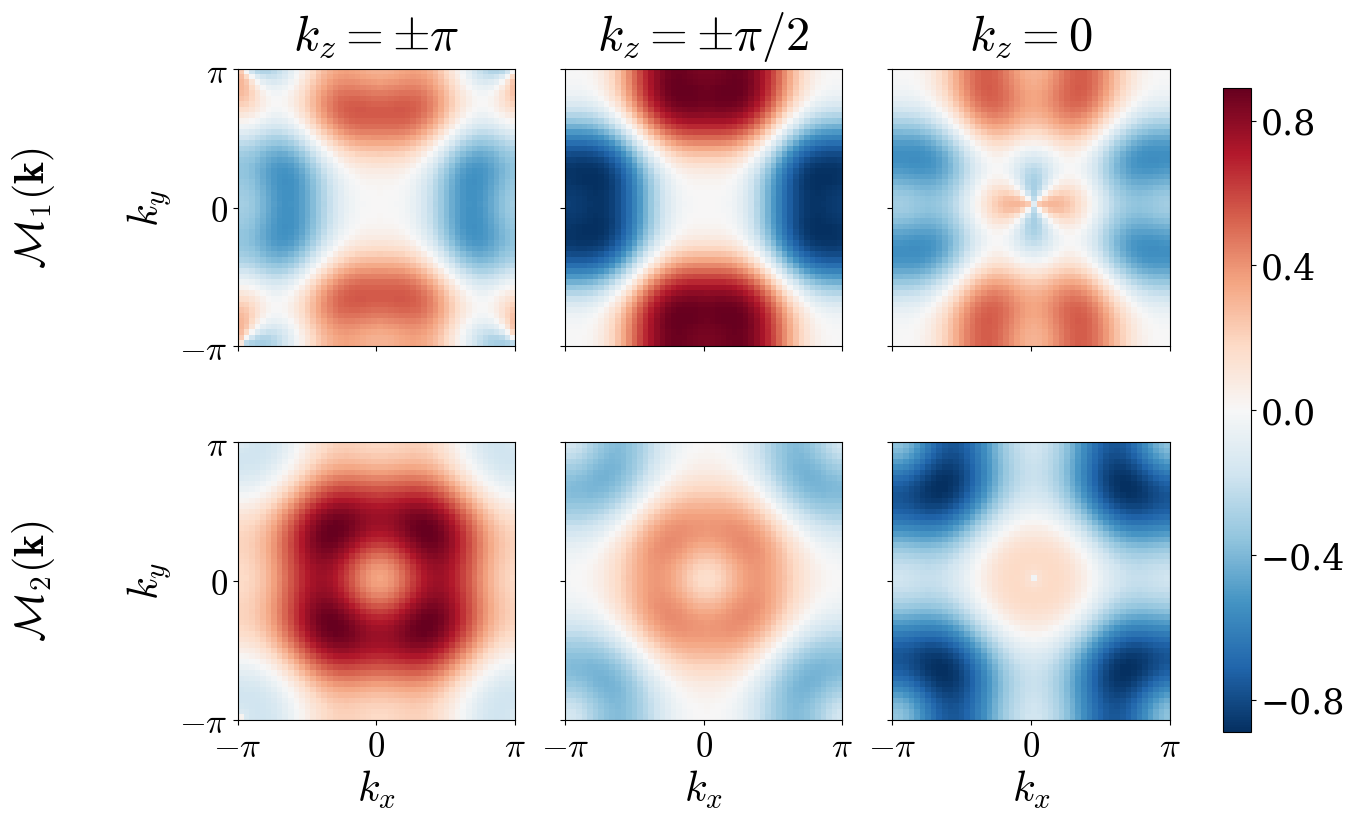}
    \caption{Plots of scattering matrix $\mathcal{M}_1(\mathbf{k})$ (top) and $\mathcal{M}_2(\mathbf{k})$ (bottom). The two functions form the two-dimensional irreducible representation $E$ of the octahedral group. Each panel shows momentum space cuts at fixed $k_z=\pm\pi, \pm\pi/2, 0$. The number of points in momentum space is 50 in each direction.}
    \label{fig:states_combined}
\end{figure}

\section{\label{sec:pairing}Pairing from Feshbach resonance}

\subsection{Scattering matrix}

The open channel, consisting of two polarons with momenta $\mathbf{k}$ and $-\mathbf{k}$, couples to the two degenerate closed channels ($\alpha=1,2$) associated with the degenerate bipolaron states. The corresponding coupling strengths are obtained from the scattering matrices $\mathcal{M}_{\alpha} (\mathbf{k})$, defined in Eq.~\ref{eq:Mform}. We compute them within the truncated basis framework and plot our results in Fig.~\ref{fig:states_combined}. For $\mathcal{M}_1(\mathbf{k})$ we find nodes at $k_y=\pm k_x$, and $\pi/2$ rotations around the $k_z$-axis yield an eigenvalue of $-1$. On the other hand, for $\mathcal{M}_2(\mathbf{k})$ there are nodes at $4k_z^{2} = k_x^{2} + k_y^{2}$, and $\pi/2$ rotations around the $k_z$-axis leave $\mathcal{M}_2(\mathbf{k})$ invariant. Therefore the functions $\mathcal{M}_1(\mathbf{k})$ and $\mathcal{M}_2(\mathbf{k})$ furnish the two basis functions of the irreducible representation $E$ of the octahedral group, inherited from the twofold degenerate bipolaron state.

\subsection{Critical Temperature}

We begin from the linearized gap equation for finite-temperature, which is valid in regimes near the critical temperature where the gap function is small~\cite{sigrist1991},
\begin{equation} \label{gapeq}
    \Delta_{\mathbf{k}} = \sum_{\mathbf{k'}} -V_{\mathbf{k}, \mathbf{k}'} \frac{\tanh (\beta \xi_{\mathbf{k'}}/2)}{2\xi_{\mathbf{k'}}} \Delta_{\mathbf{k'}},
\end{equation}
where $\xi_k=\epsilon_{\mathbf{k}}^{\text{sc}}-\mu$ and $\beta = 1/(k_B T)$ is the inverse temperature. 

The appropriate ansatz for the gap is obtained by substituting Eq.~\ref{eq:potential_expanded} into Eq.~\ref{gapeq}. Summing over $\mathbf{k'}$ and comparing the resulting $\mathbf{k}$-dependent terms yields the following,
\begin{equation} \label{eq:ansatz}
    \Delta_{\mathbf{k}} = (c_1 +  \frac{c_2}{2\epsilon_{\mathbf{k}}^{\text{sc}}-E_0})(\psi_1 \mathcal{M}_1(\mathbf{k}) + \psi_2 \mathcal{M}_2(\mathbf{k})).
\end{equation}
where $c_1$ and $c_2$ are real coefficients determined self-consistently by substituting Eq.~\ref{eq:ansatz} into Eq.~\ref{gapeq}. The parameters $\psi_1$ and $\psi_2$ are the complex order parameter components associated with the irreducible representation $E$ of the octahedral group. $E_0$ denotes the energy of the bipolaron at momentum $(\pi,\pi,\pi)$. Eq.~\ref{eq:ansatz} shows that the gap function $\Delta_{\mathbf{k}}$ inherits the symmetry properties of the scattering matrices $\mathcal{M}_{\alpha} (\mathbf{k})$. 

\begin{figure}[t] 
    \centering
    \includegraphics[width=0.95\columnwidth]{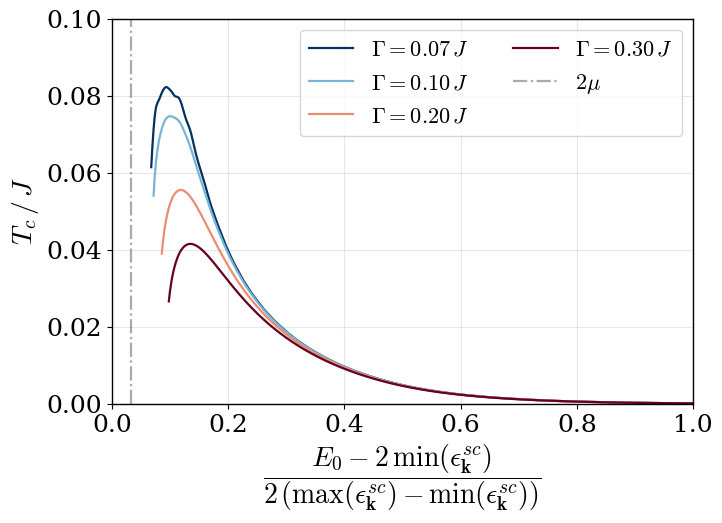}
    \caption{Critical temperature $T_c$, measured in units of $J$, plotted against the detuning from resonance $E_0-2 \min \epsilon_{\mathbf{k}}^{\text{sc}}$ for a fixed chemical potential($\mu=\min(\epsilon_{\mathbf{k}}^{\text{sc}})+0.03t$), for various inverse lifetimes $\Gamma$ of the bound state. The x-axis is scaled so that the bound state energy $E_0$ is between the minimum and maximum values of $2\epsilon_{\mathbf{k}}^{\text{sc}}$. The maximum value of the critical temperature reaches $\simeq 0.08 J$ for the inverse lifetime $\Gamma=0.1J$.}
    \label{fig:lifetime}
\end{figure}

If the bound state energy $E_{0}$ satisfies $\min(2 \epsilon_{\mathbf{k}}^{\text{sc}})<E_{0} < \max(2 \epsilon_{\mathbf{k}}^{\text{sc}})$, summing over the entire Brillouin zone leads to divergences because of zeros in the denominator. To address this, we take into account the finite lifetime $\Gamma$ of the metastable bipolaron branch, by replacing $E_0 \to E_0 - i \Gamma$. Field theoretic estimates yield inverse lifetimes on the order of $0.1J$, shown in Appendix~\ref{app:lifetime}. 

Eq.~\ref{gapeq} is solved with the gap ansatz in Eq.~\ref{eq:ansatz}, accounting for the finite lifetime of bipolarons. The equations for $\psi_1$ and $\psi_2$ decouple and can be solved independently; both yield the same critical temperature. The resulting critical temperature is shown for different values of $E_0$ in Fig.~\ref{fig:lifetime}. The different curves correspond to the different lifetimes of the bipolaron. The effective interaction between two polarons scales inversely with the energy difference between the two channels. The maximum value of the critical temperature exhibits an inversely proportional behavior to the inverse lifetime. Introducing a finite lifetime of the bipolarons has a noticeable effect near resonance, and for off-resonant values of the bound state energy the values of critical temperature remain largely unchanged. 

\begin{figure}[t]
    \centering
    \includegraphics[width=0.95\columnwidth]{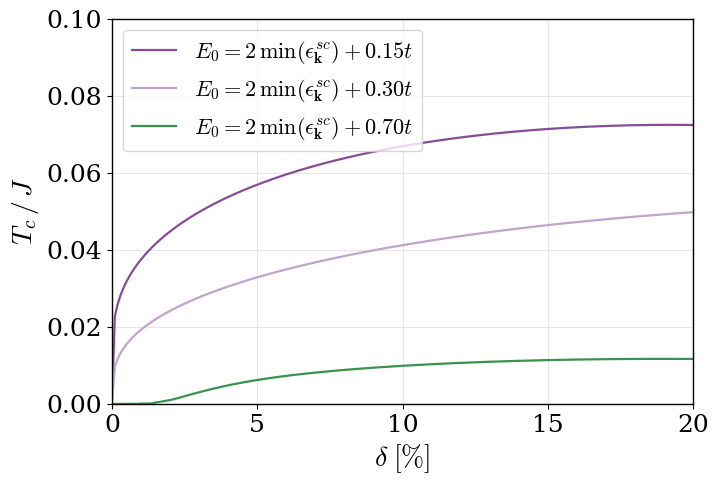}
    \caption{Relationship between doping $\delta$ and critical temperature $T_c$, measured in units of $J$. Both channels, $\mathcal{M}_1(\mathbf{k})$ and $\mathcal{M}_2(\mathbf{k})$ yield the same critical temperature. At larger doping levels, when the AFM order is destroyed, our theory is no longer applicable.}
    \label{fig:doping}
\end{figure}

For various chemical potentials, we compute the critical temperatures and doping fractions to obtain the critical temperature as a function of doping, as shown in Fig.~\ref{fig:doping}. We keep $E_0$ fixed. The critical temperature initially increases with doping and eventually reaches a plateau. The distinct superconducting channels both yield the same critical temperature. We conclude that critical superconducting temperatures on the order of $T_c\simeq0.1J$ can be reached if the system is tuned close to the emergent Feshbach resonance. 

\subsection{Weyl superconductivity}

From symmetry arguments alone, one can deduce the stable superconducting phases once it is known which irreducible representation the order parameter follows. In our case the system has a two component complex order parameter $(\psi_1, \psi_2)$, with each component corresponding to the amplitude for that channel \cite{gao2026}. The following equation, which describes a free energy expansion near the phase transition of an order parameter following the irreducible representation $E$ of the $O$ group, is taken from Ref.~\cite{gao2026}, and can be obtained using the Clebsch-Gordon formalism~\cite{sigrist1991},
\begin{equation}
\begin{aligned}
F &= \alpha \left(|\psi_1|^2 + |\psi_2|^2\right)
   + \beta_1 \left(|\psi_1|^2 + |\psi_2|^2\right)^2 \\
  &\quad + \beta_2 \left|\psi_1^{*}\psi_2 - \psi_2^{*}\psi_1\right|^2
   + \cdots.
\end{aligned}
\label{eq:GLeq}
\end{equation}

\subsubsection{Time-reversal symmetry breaking}

As discussed in Appendix~\ref{app:symmetry}, a superconductor with a two-dimensional order parameter in 3D with octahedral symmetry only supports two stable superconducting states~\cite{sigrist1991}, the time-reversal symmetry breaking (TRSB) state and the nematic state breaking the underlying crystal symmetry $O$. Which state the system is in is decided by the sign of $\beta_2$, the fourth order mixing term coefficient of the Ginzburg-Landau free energy expansion (see Eq.~\ref{eq:GLeq}). From our microscopic model, the Ginzburg-Landau coefficients can be calculated perturbatively using the Hubbard-Stratonovich transformation~\cite{altland2023}. A derivation is provided in Appendix~\ref{app:ginzburg}. We obtain:
\begin{equation} \label{eq:beta}
    \beta_2=-\sum_{\mathbf{k}} \gamma_{1}^2(\mathbf{k})\gamma_{2}^2(\mathbf{k})\left[ \frac{\tanh(\xi_{\mathbf{k}}/2T)}{4\xi_{\mathbf{k}}^3} - \frac{\text{sech}^2(\xi_{\mathbf{k}}/2T)}{8T\xi_{\mathbf{k}}^2} \right].
\end{equation}
Here $\gamma_{1\mathbf{k}}$ and $\gamma_{2\mathbf{k}}$ denote the normalized form factors obtained from $\mathcal{M}_1$ and $\mathcal{M}_2$ respectively, via the relation $\mathcal{M}_{\alpha}(\mathbf{k})\mathcal{M}_{\alpha}(\mathbf{k}')=g \gamma_{\alpha}(\mathbf{k})\gamma_{\alpha}(\mathbf{k}')$, where $g$ is a constant. Substituting the numerical values for the form factors $\gamma_{1\mathbf{k}}, \gamma_{2\mathbf{k}}$ and energy $\xi_k$, one finds $\beta_2\approx-0.002$ for the temperatures below $T_c$ and $\mu$ values corresponding to low filling. $\beta_2$ has a negative value, implying TRSB superconductivity with $d_{x^2-y^2}+id_{z^2}$ pairing. To summarize, we predict that the superconducting phase in the 3D repulsive Hubbard model not only breaks the $U(1)$ gauge symmetry but also time-reversal symmetry, and preserves the octahedral crystal symmetry of the Hamiltonian.

\subsubsection{Weyl points}

The superconducting gap for a $d_{x^2-y^2}+id_{z^2}$ order parameter vanishes along nodal lines in momentum space. Nodal lines of the gap coincide with the nodal lines of the scattering matrices $\mathcal{M}_{\alpha} (\mathbf{k})$ (see Fig.~\ref{fig:states_combined}), therefore the nodes lie on the $(0,0,0)-(\pi,\pi,\pi)$ diagonal of the Brillouin zone and its symmetry equivalents. Since the Fermi pockets are centered around $(\pi/2,\pi/2,\pi/2)$ and equivalent points, for a $d_{x^2-y^2}+id_{z^2}$ order parameter there are always two Weyl points per octant of the crystal Brillouin zone (see Fig.~\ref{fig:overview}). Where these nodal lines intersect the Fermi pockets, they produce isolated point nodes in the Bogoliubov quasiparticle spectrum, around which the low-energy quasiparticle excitation disperses linearly.

\section{\label{sec:conclusion}Discussion and Outlook}

We predict that the superconducting phase of the three-dimensional Fermi-Hubbard model in the strongly interacting regime spontaneously breaks time-reversal symmetry, resulting in a $d_{x^2-y^2}+id_{z^2}$ pairing symmetry and realizing a Weyl superconducting phase. This result was obtained by analyzing polarons and bipolarons in cubic lattices with AFM order. A Fermi liquid of polarons forms at low doping, and the polarons scatter with each other to form virtual bipolaronic bound states. This scattering results in an effective interaction between the polarons, and we have deduced the associated pairing symmetry. The pairing symmetry was shown to be $d$-wave, which means that it has two nodes passing through the origin. Compared to the 2D square lattice, the extra spatial dimension allows for two $d$-wave pairing channels with the same critical temperature. In our analysis, we have neglected intra-pocket pairing and focused instead on inter-pocket pairing mediated by coupling to the $\mathrm{R}$-point (see Fig.~\ref{fig:bipolaron}) of the bipolaron dispersion, as this corresponds to the lowest-energy state accessible. In addition, since we only consider singlet pairing, coupling to the $\Gamma$-point is irrelevant because it gives rise to an odd spatial symmetry.

Our mechanism for superconductivity is based on an emergent Feshbach resonance in the Hubbard model. While our calculations based on the truncated basis approach suggest general proximity to the Feshbach resonance, determining the precise location of the resonance will require dedicated large-scale numerical simulations or experiments. Additional tuning parameters such as next-to-nearest neighbor hopping may be required to reach near-resonant interactions, which provide valuable tuning parameters that can be explored in future works. 

Cold atom experiments cannot yet be prepared at low enough temperatures to observe superconductivity, but additional entropy redistribution schemes might make observation possible \cite{mazurenko2017}. Another interesting future direction is the exploration of superconducting surface states associated with the Weyl points we predict.

\begin{acknowledgments}
We thank Linus Hein, Annika Böhler, Lukas Homeier, Hannah Lange, Annabelle Bohrdt, Eugen Dizer, and Eugene Demler for fruitful discussions. T.K. was supported by a scholarship of the German Academic Exchange Service (DAAD). This project has received funding from the European Research Council (ERC) under the European Union’s Horizon 2020 research and innovation programm (Grant Agreement no 948141) — ERC Starting Grant SimUcQuam, and by the Deutsche Forschungsgemeinschaft (DFG, German Research Foundation) under Germany's Excellence Strategy -- EXC-2111 -- 390814868.

\end{acknowledgments}

\appendix

\section{Lee-Low-Pines transformation} \label{app:LLP}

We closely follow Ref.~\cite{bermes2024} and show the results of the Lee-Low-Pines transformation. We introduce a hole momentum operator $\hat{\mathbf{X}}_h = \sum_{\mathbf{j}} \mathbf{r}_{\mathbf{j}} \left( 1 - \sum_{\sigma} \hat{c}^{\dagger}_{\mathbf{j}\sigma} \hat{c}_{\mathbf{j}\sigma} \right)$ and a total momentum operator of the fermions $\hat{\mathbf{Q}}_c = \sum_{\mathbf{k}\sigma} \mathbf{k} \hat{c}^\dagger_{\mathbf{k}\sigma}\hat{c}_{\mathbf{k}\sigma}$ to construct a unitary operation, which shifts the reference frame to one comoving with the hole,

\begin{equation}
    \hat{U}_{\mathrm{LLP}} = \exp(i \hat{\mathbf{X}}_h \hat{\mathbf{Q}}_c).
\end{equation}

Applying this unitary transformation along with a Fourier transform block diagonalizes the Hamiltonian in the momentum basis,

\begin{equation}
    \hat{\mathcal{H}}= \sum_{\mathbf{k}\sigma} \hat{\mathcal{H}}(\mathbf{k})\hat{c}^\dagger_{\mathbf{k}\sigma}\hat{c}_{\mathbf{k}\sigma},
\end{equation}
where $\hat{\mathcal{H}}(\mathbf{k})$ is given by

\begin{equation}
\begin{aligned}
\hat{\mathcal{H}}(\mathbf{k})
&= \langle \mathbf{k} | \hat{U}_{\mathrm{LLP}}^\dagger \hat{\mathcal{H}}_{\mathrm{eff}}\hat{U}_{\mathrm{LLP}} | \mathbf{k} \rangle \\
&= t \sum_{\delta\sigma} \bigl(e^{i(\hat{\mathbf{Q}}_c -\mathbf{k})\delta} \hat{c}^\dagger_{0,\sigma}\hat{c}_{\delta,\sigma}+\text{h.c.}\bigr) \\
&\quad + J_z \sum_{\langle \mathbf{i},\mathbf{j} \rangle}
\left( \hat{S}^z_{\mathbf{i}} \hat{S}^z_{\mathbf{j}} - \frac{1}{4}\hat{n}_{\mathbf{i}} \hat{n}_{\mathbf{j}} \right) \\
&\quad + \frac{J_\perp}{2} \sum_{\langle \mathbf{i},\mathbf{j} \rangle}
\left( \hat{S}^+_{\mathbf{i}} \hat{S}^-_{\mathbf{j}} + \text{h.c.} \right).
\end{aligned}
\end{equation}

\section{Group theoretical analysis} \label{app:group}

The ground state of the bipolaron is twofold degenerate at the $(\pi, \pi, \pi)$ point. Since the octahedral group possesses a unique two-dimensional irreducible representation $E$~\cite{dresselhaus2008}, it is natural to identify this degeneracy with the representation $E$. This is supported by the observation that applying the symmetry operations of the octahedral group to the degenerate states yields eigenvalues whose sums reproduce the character values of the representation $E$, as shown in Table~\ref{table:group}.

\begin{table}[htbp]
\centering
\renewcommand{\arraystretch}{1.5}
\resizebox{\columnwidth}{!}{
\begin{tabular}{l|cc|c|c}
\hline
\textbf{Rotation} & \textbf{Eigenvalue 1} & \textbf{Eigenvalue 2} & \textbf{Sum} & \textbf{Character ($\chi$)} \\ \hline
$C_3$ & $-0.4956 + 0.8584i$ & $-0.5044 - 0.8736i$ & $-1.0000 - 0.0152i$ & $-1$ \\
$C_2$ & $1.0042$ & $0.9958$ & $2.0000$ & $2$ \\
$C_4$ & $1.0153$ & $-0.9846$ & $0.0307$ & $0$ \\ \hline
\end{tabular}%
}
\caption{Numerical eigenvalues of different rotation operators of the two degenerate ground states of two holes at momentum $(\pi, \pi, \pi)$. The sum of the eigenvalues of each rotation operator equal the character of the irreducible representation $E$ of the octahedral group. The character values are taken from Ref.~\cite{dresselhaus2008}}
\label{table:group}
\end{table}

\section{\label{app:lifetime}Lifetime of state}

When accounting for the finite lifetime of the bipolarons, Eq.~\ref{eq:potential_expanded} assumes the following form,
\begin{equation}
\begin{split}
V_{\mathbf{k},\mathbf{k}'} &= \frac{1}{2V}\sum_{\alpha=1}^{2}
\mathcal{M}_{\alpha}(\mathbf{k})\mathcal{M}_{\alpha}(\mathbf{k}')\\
&\quad\times\left[
\frac{1}{2\epsilon_{\mathbf{k}}^{\mathrm{sc}}-E_0+i\Gamma}
+\frac{1}{2\epsilon_{\mathbf{k}'}^{\mathrm{sc}}-E_0+i\Gamma}
\right]
\end{split}
\label{eq:gamma}
\end{equation}
We aim to obtain an estimate for the magnitude of the inverse lifetime $\Gamma$. Ref.~\cite{bermes2026} derives a formula for the self-energy of the bipolaron state at zero doping in a 2D AFM system. Extending this equation into 3D, and accounting for the extra coupling term, the self-energy can be written as the following,
\begin{equation}
    \Pi(\mathbf{k},\nu) = \int \frac{d^3p}{(2\pi)^3} \frac{|M_{\mathbf{p}}^{+1}|^2+|M_{\mathbf{p}}^{-1}|^2}{\nu - \epsilon_{\mathbf{p}}^{\mathrm{sc}} - \epsilon_{\mathbf{k-p}}^{\mathrm{sc}} + i\eta}.
\end{equation}
The imaginary part of this is found to be
\begin{equation}
\begin{aligned}
\operatorname{Im}\Pi(\mathbf{k},\nu)
={}&-\int \frac{d^3p}{(2\pi)^3}\,
\Bigl(
    \lvert M_{\mathbf p}^{+1}\rvert^2
   +\lvert M_{\mathbf p}^{-1}\rvert^2
\Bigr)
\\[-2pt]
&\times
\frac{\eta}{
\bigl(\nu-\epsilon_{\mathbf p}^{\mathrm{sc}}
-\epsilon_{\mathbf{k-p}}^{\mathrm{sc}}\bigr)^2+\eta^2
}.
\end{aligned}
\end{equation}

Since the momentum grid is finite, we cannot take the limit $\eta \to 0^+$ and evaluate the imaginary part of the self-energy using the Dirac identity. Instead, take a value of $\eta$ in the regime in which we obtain smoothening but does not smear the full spectrum. 

At $(\pi,\pi,\pi)$, one must find the energy value $E^*$ that satisfies $E^* - E_0 - \mathrm{Re}\,\Pi((\pi,\pi,\pi),E^*) = 0$, where $E_0$ is the energy of the bipolaron at momentum $(\pi,\pi,\pi)$. Assuming that $E_0$ has a value of $0.1t$, one can sum over the momenta and obtain the following quantity,
\begin{equation}
    \Gamma = - \operatorname{Im}(\Pi((\pi,\pi,\pi),E^*)).
\end{equation}
The inverse lifetime $\Gamma$ is found to have values on the magnitude of $0.1J$.

\section{\label{app:symmetry}More about Ginzburg-Landau expansion}

We show Eq.~\ref{eq:GLeq} again for clarity,
\begin{equation}
\begin{aligned}
F &= \alpha \left(|\psi_1|^2 + |\psi_2|^2\right)
   + \beta_1 \left(|\psi_1|^2 + |\psi_2|^2\right)^2 \\
  &\quad + \beta_2 \left|\psi_1^{*}\psi_2 - \psi_2^{*}\psi_1\right|^2
   + \cdots.
\end{aligned}
\end{equation}
The first term controls the superconducting phase transition \cite{altland2023}. For $\alpha>0$, $\psi_1$ and $\psi_2$ are 0 to minimize the free energy, and this corresponds to the regime above the critical temperature. For $\alpha<0$, the order parameter acquires a non-zero value, corresponding to the symmetry-broken regime below the critical temperature. Here one must also consider the fourth order terms, since otherwise the order parameter can diverge. To prevent this, $\beta_1>0$ must be satisfied for stability, and $\beta_2$ can be positive or negative. $\beta_2>0$ leads to $\left|\psi_1^{*}\psi_2 - \psi_2^{*}\psi_1\right|=0$, and $(\psi_1,\psi_2)= \Delta(\cos\theta,\sin \theta)$ satisfies this condition, where $\Delta$ can be chosen to be real and $\theta=\tan (\psi_2/\psi_1)$. This corresponds to the nematic phase, where the order parameter breaks rotational symmetry. On the other hand, $\beta_2<0$ makes the system maximize the term $\left|\psi_1^{*}\psi_2 - \psi_2^{*}\psi_1\right|^2$, and therefore $\psi_1=i\psi_2$ or $\psi_1=-i\psi_2$ are possible solutions. The order parameter then breaks time reversal symmetry, corresponding to TRSB superconductivity.

\section{\label{app:ginzburg}Calculation of Ginzburg-Landau coefficients}

The Ginzburg-Landau coefficients can be calculated from the dispersion relation of polarons shown in Fig.~\ref{fig:singlehole_combined_aligned} and the scattering matrices shown in Fig.~\ref{fig:states_combined}. We employ a similar calculation as Ref.~\cite{altland2023}, but with a two-dimensional order parameter. We will use the approximation $\Delta E=2\epsilon_{\mathbf{k}}^{\text{sc}}-E_{0}=2\epsilon_{\mathbf{k'}}^{\text{sc}}-E_{0}$ to simplify the interaction term. This is justified because the coefficients depend on the sum of quantities over the entire Brillouin zone, multiplied by a weight peaked at the Fermi surface,
\begin{equation}
    V_{\mathbf{k},\mathbf{k}'} = -\frac{g}{V\Delta E}\left[
    \gamma_{1}(\mathbf{k})\gamma_{1}(\mathbf{k}')
    + \gamma_{2}(\mathbf{k})\gamma_{2}(\mathbf{k}')
    \right].
\end{equation}

Here, the values $\gamma_\alpha(\mathbf{k})$ with $\alpha=1,2$ are normalized such that $\sum_\mathbf{k} \gamma_{\alpha}(\mathbf{k})^2 = 1$, with $g$ chosen to satisfy $\mathcal{M}_{\alpha}(\mathbf{k})\mathcal{M}_{\alpha}(\mathbf{k}')=g \gamma_{\alpha}(\mathbf{k})\gamma_{\alpha}(\mathbf{k}')$. $V$ is the volume of the system. Since there are two interaction terms, we introduce two bosonic fields $\psi_1$, $\psi_2$ using the Hubbard-Stratonovich transformation to decouple the interaction term. The two-dimensional order parameter can be written as $\Delta_{\mathbf{k}} = \psi_1\, \gamma_{1}(\mathbf{k}) + \psi_2\, \gamma_{2}(\mathbf{k})$, and $\psi_1$ and $\psi_2$ can be interpreted as complex amplitudes of the two channels. The path integral for the partition function $\mathcal{Z}$ after integrating out the polaron fermion fields yields a path integral of the $\psi_1$, $\psi_2$ boson fields,
\begin{equation}
    \mathcal{Z} = \int D\psi_1 D\psi_2 \exp\left( -S_E \right),
\end{equation}
where the effective action is
\begin{equation}
    S_E =\int d\tau \frac{V}{g_0}\left(|\psi_1|^2 + |\psi_2|^2\right) - \sum_{\mathbf{k}} \operatorname{tr} \ln \hat{\mathcal{G}}_{\mathbf{k}}^{-1}.
\end{equation}
The inverse Gor'kov Green function $\hat{\mathcal{G}}_k^{-1}$ in Nambu space is
\begin{equation}
    \hat{\mathcal{G}}_{\mathbf{k}}^{-1} = \begin{pmatrix} 
    i\omega_n - \xi_{\mathbf{k}} & \Delta_{\mathbf{k}} \\ 
    \bar{\Delta}_{\mathbf{k}} & i\omega_n + \xi_{\mathbf{k}} 
    \end{pmatrix},
\end{equation}
with $\omega_n$ denoting the fermionic Matsubara frequency. The corresponding Nambu spinors are defined as
\begin{equation}
    \bar{\Pi}_{\mathbf{k}}
    = (\bar{\pi}_{\mathbf{k}\uparrow} \quad \pi_{-\mathbf{k}\downarrow}),
    \quad
    \Pi_{\mathbf{k}}
    = \begin{pmatrix}
        \pi_{\mathbf{k}\uparrow} \\
        \bar{\pi}_{-\mathbf{k}\downarrow}
      \end{pmatrix}.
\end{equation}
We can split the inverse Gor'kov Green function into a non-interacting part $\hat{\mathcal{G}}_{0,\mathbf{k}}^{-1}$ and an interacting part $\hat{\Delta}_{\mathbf{k}}$ as $\hat{\mathcal{G}}_{\mathbf{k}}^{-1}=\hat{\mathcal{G}}_{0,\mathbf{k}}^{-1}+\hat{\Delta}_{\mathbf{k}}$, where 
\begin{equation}
    \hat{\mathcal{G}}_{0,\mathbf{k}}^{-1} \equiv \begin{pmatrix} 
    i\omega_n - \xi_{\mathbf{k}} & 0 \\ 
    0 & i\omega_n + \xi_{\mathbf{k}} 
    \end{pmatrix}, \quad \hat{\Delta}_{\mathbf{k}} \equiv \begin{pmatrix} 0 & \Delta_{\mathbf{k}} \\ \bar{\Delta}_{\mathbf{k}} & 0 \end{pmatrix}.
\end{equation}
Near the critical temperature, where $\hat{\Delta}_{\mathbf{k}}$ is small, it is possible to expand the logarithmic trace in the effective action. The fourth order term of this expansion is obtained using the approximation that $\psi_{1}$ and $\psi_{2}$ are uniform and static, and is shown in Eq.~\ref{eq:fourth},
\begin{equation} \label{eq:fourth}
    \begin{split}
-\frac{1}{4}\operatorname{tr}\bigl(\hat{\mathcal{G}}_{0,\mathbf{k}}\hat{\Delta}_\mathbf{k}\bigr)^4
&= -\frac{1}{2L^d}\sum_{\mathbf{k}}
   \bigl|\psi_1\, \gamma_{1}(\mathbf{k}) + \psi_2\, \gamma_{2}(\mathbf{k})|^4 \\
&\quad\times\left[\frac{\tanh(\xi_{\mathbf{k}}/2T)}{4\xi_{\mathbf{k}}^3}
   -\frac{\operatorname{sech}^2(\xi_{\mathbf{k}}/2T)}{8T\xi_{\mathbf{k}}^2}\right].
\end{split}
\end{equation}

Expanding this term and summing over the momenta yields the fourth order Ginzburg-Landau coefficients. Note that terms such as $\gamma_1^3\gamma_2$ and $\gamma_1\gamma_2^3$ cancel out after the summation over the momenta due to orthogonality. Comparing the terms in the above equation to Eq.~\ref{eq:GLeq}, we obtain the equation for $\beta_2$, shown in Eq.~\ref{eq:beta}. Because the expression in brackets is strictly positive, $\beta_2$ is necessarily negative in a weak-coupling theory, thereby ensuring that the TRSB state is always preferred.

\end{document}